\documentclass[conference,letterpaper,10pt]{IEEEtran}

\usepackage[latin1]{inputenc}
\usepackage{times,amsmath}
\usepackage{amsfonts}
\usepackage{pstool}
\usepackage{multirow}
\usepackage{enumerate}
\usepackage{graphicx}
\usepackage{MnSymbol}
\usepackage{stfloats}
\usepackage[table]{xcolor}
\usepackage[square, comma, sort&compress, numbers]{natbib}
\usepackage{nohyperref}
\usepackage{algorithm,algorithmic}
\usepackage{bigints}
\usepackage{amsmath}
\usepackage{subfig}

\begin{document}

\title{Design of an Efficient Single-Stage and 2-Stages Class-E Power Amplifier (2.4GHz) for Internet-of-Things}

\author{
\IEEEauthorblockN{ Ayyaz Ali\IEEEauthorrefmark{1}, Syed Waqas Haider Shah\IEEEauthorrefmark{2}, Khalid Iqbal\IEEEauthorrefmark{3} }
\IEEEauthorblockA{\IEEEauthorrefmark{1}Department of Electrical Engineering, Army Public College of Management \& Sciences, Rawalpindi (46000), Pakistan\\\IEEEauthorrefmark{2}Department of Electrical Engineering, Information Technology University, Lahore (54000), Pakistan\\\IEEEauthorrefmark{3} National University of Science and Technology, Islamabad (46000), Pakistan\\\ ayyazali21@yahoo.com, waqas.haider@itu.edu.pk, kiqbal@ceme.nust.edu.pk}
}
\maketitle
\begin{abstract}
In this work, the designs of a single-stage and 2-stage 2.4 GHz power amplifier (PA) are presented. The proposed PAs have been designed to provide high gain and improved efficiency using harmonic suppression and optimized impedance matching techniques. There are two harmonic suppression circuits, each stage of the PA consists of 2 capacitors and 2 inductors, which will help to suppress the harmonic frequency for 2.4 GHz. These suppression circuits will help to enhance the overall efficiency of the PAs. Both the PAs are provided with a VCC supply of $4.2 V$. Input and output impedances are matched to 50 ohms. Simulation and experimental results are presented, where the simulated gain and power added efficiency (PAE) for single stage PA are $17.58 dB$ and $53\%$, respectively. While the experimental gain and PAE are $16.7 dB$ and $49.5\%$, respectively. On the other hand, for 2-stages PA, simulated gain comes out to be $34.6 dB$ and PAE is $55\%$, while the experimental gain and PAE are $30.5 dB$ and $53.1\%$, respectively. The final design is being fabricated on the Taconic printed circuit board (PCB) with a thickness of $0.79 mm$ and dielectric constant value of 3.2 and its dimensions are $4.6 cm \times 3.4 cm$ for single stage and $5.9 cm \times 3.6 cm$ for 2-stages PA.
\end{abstract}
\IEEEpeerreviewmaketitle
\begin{IEEEkeywords}
2.4 GHz power amplifier, class-E power amplifier, power added efficiency (PAE), Internet-of-Things
\end{IEEEkeywords}
\section{Introduction}
There has been a lot of development in wireless communication in recent years. Especially with the advent of Internet-of-Things (IoT), need for low cost and low power wireless communication systems are increasing rapidly with small size and enhanced efficiency. One of the most important parts of any wireless communication network is a power amplifier (PA) present at the transmit station \cite{shah2016adaptive}. Most of the IoT applications requires battery operated devices such as home and factory automation, smart cities and connected agriculture. These devices are designed such that they consume lesser and lesser power for their operation. Major consumption of power of a device is associated with its transmitter section, in which further mostly power is consumed by the PA of that transmitter. Such PAs need to be efficient enough so that they may use little battery power and provide enough gain which will enhance the power of an input signal. In any PA, two main characteristics which need the most consideration are the efficiency and linearity \cite{van2009compact}. When designing a PA, the selection of the right circuit, class topology and semiconductor greatly affects the linearity and efficiency. PAs can basically be classified as linear and nonlinear. The efficiency of nonlinear PA is much greater when compared to linear PA \cite{jie20102}. Out of different efficient classes (Class D, E, and F) of PA, Class-E is a suitable choice for the design of an efficient PA with switch mode input circuitry.
\begin{table}[]
\centering
\caption{Comparison of Linearity and Efficiency between different Classes of PA}
\label{my-label}
\begin{tabular}{|c|c|c|c|}
\hline
\textbf{Class} & \textbf{Ideal Efficiency} & \textbf{Linearity} & \textbf{Practical Efficiency} \\ \hline
Class-A        & 50 \%                     & Good               & 35\%                          \\ \hline
Class-AB       & 50\% - 78.5\%             & Good               & 45\%                          \\ \hline
Class-B        & 78.5\%                    & Moderate           & 49\%                          \\ \hline
Class-C        & 78.5\% - 100\%            & Poor               & 55\%                          \\ \hline
Class-E        & 100\%                     & Poor               & 62\%                          \\ \hline
Class-F        & 100\%                     & Poor               & 80\%                          \\ \hline
\end{tabular}
\end{table}
\begin{figure}[ht]
\begin{center}
	\includegraphics[width=2.7in]{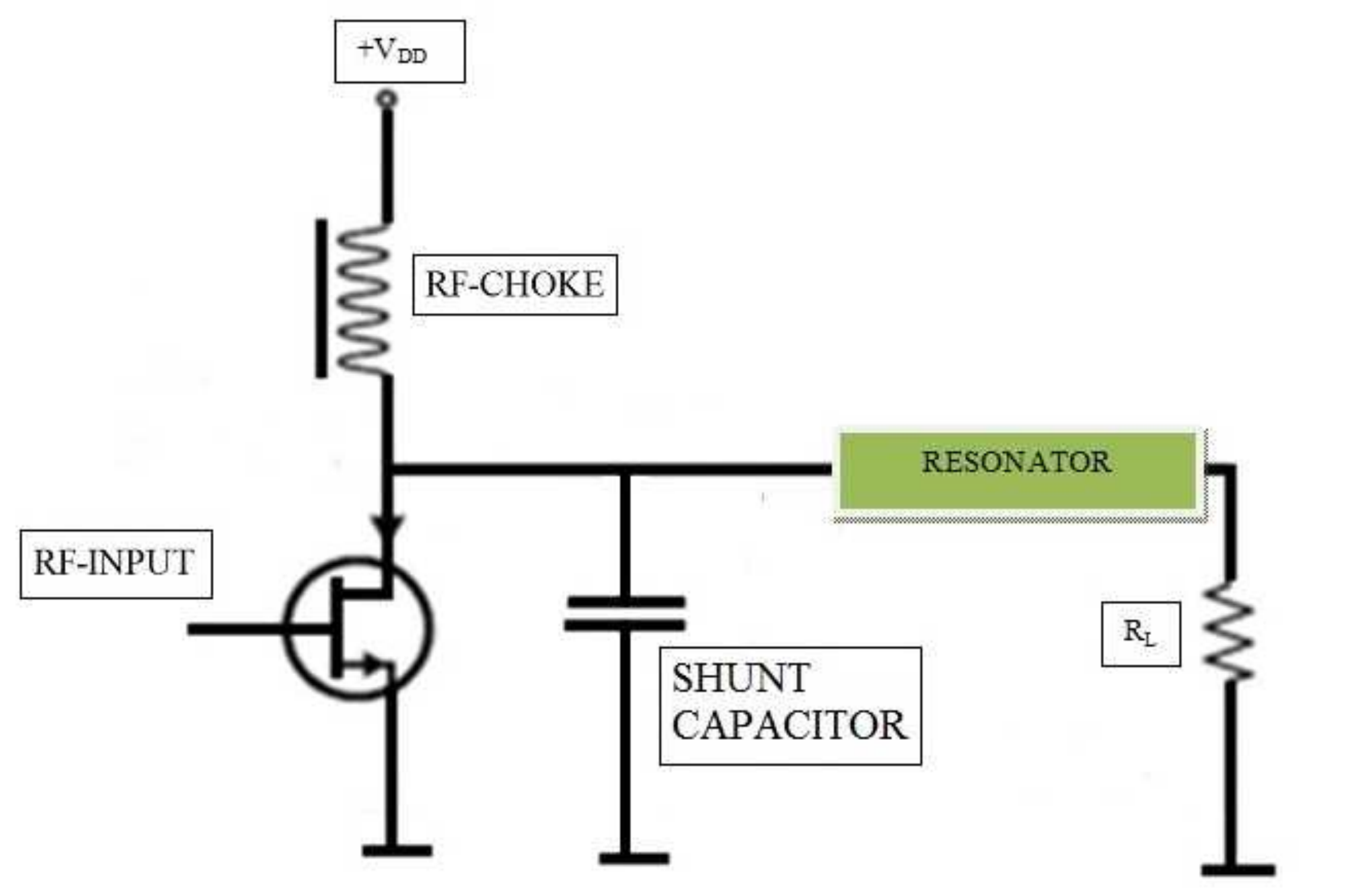}
\caption{Basic design of Class-E PA.}
\label{fig:sysmodel}
\end{center}
\end{figure}
A comparison of linearity and efficiency between different classes of PAs are shown in Table 1 from where it can be seen that a Class-E topology for the design of an efficient PA would be a good choice. From \cite{ostrovskyy2012performance} the selection of the topology is highly dependent upon the radio frequency (RF) choke size. If this size is to be finite in nature then the design calculation formula for Class-E PA will not be well-suited. Keeping the frequency and output power fixed, the Class-E PA have a power loss of 2.3 less as compared to Class-B or Class C amplifiers \cite{raab1977idealized}.

A simple PA is shown in Fig. 1 with Class-E configuration. Generally, there are two major types of amplifiers which are zero voltage switching (ZVS) and zero current switching (ZCS) \cite{meshkin2010novel}. The transistor in this class works as a switch which unlike in other linear amplifiers works simply as a current source. In Fig. 1 there is a resonator, a shunt capacitor and a RF choke which are used to change the phase of output current and voltage. The RF choke helps to increase the output power of the amplifier, without which, it becomes difficult to achieve due to the limitations of the shunt capacitor \cite{thian20102}. When a signal is applied at the input of transistor it switches according to that signal. The resonator circuitry at the output impedance matching circuit is configured such that it only allows the first harmonic frequency of operation to be available at the load. As stated by SOKAL in soft switching there is no overlapping of current and voltage waveform in ideal Class-E amplifier so no power is dissipated and its efficiency would be 100\%.
\begin{figure}[ht]
\begin{center}
	\includegraphics[width=3in]{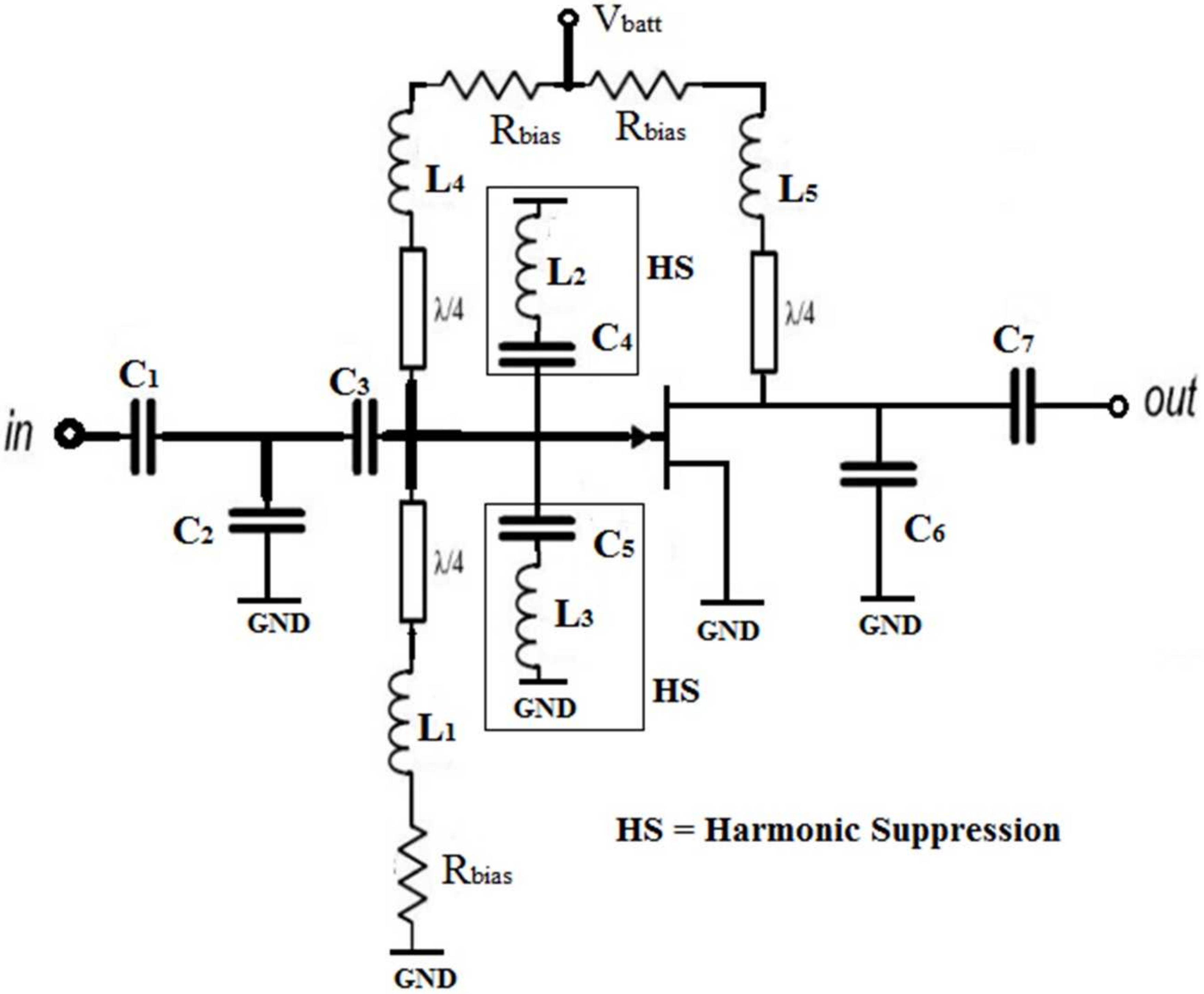}
\caption{Circuit diagram of Class-E PA with Harmonic Suppression circuit.}
\label{fig:ectheta}
\end{center}
\end{figure}

The Capacitor $C$ and inductor $L$ together helps to suppress the harmonic frequencies at the load impedance matching network. The shunt capacitor helps to reduce the power losses of switching, unlike any other topology. Since the shunt capacitor at the output introduces delays at the rise time of turning ON and OFF of the transistor, this enables the soft switching \cite{wu2011design,ali2017design}. This soft switching property makes sure that the voltage and current at the switching time never become nonzero simultaneously. In an ideal Class-E P.A when the transistor is switched OFF then the voltage is maximum and current is minimum and when the transistor is switched ON the voltage drops down to its minimum value and current reaches to a maximum value and this helps in reducing the power dissipation while switching. To attain this characteristic, the frequency of operation must exist in the range of $R1 < R < R2$, where expressions for $R_1$, $R_2$, and $C_{ap}$ are given as follows,
\begin{equation}
C_{ap} = \frac{C_1 \times C_2}{C_1 + C_2}
\end{equation}
\begin{equation}
R_1 = \frac{1}{2\pi \sqrt{L_s \times C_1}}
\end{equation}
\begin{equation}
R_2 = \frac{1}{2\pi \sqrt{\frac{L_s \times C_1 \times C_2}{C_1 + C_2}}}
\end{equation}
\subsection{Contribution \& Organization}
This work has the following contributions:
\begin{itemize}
  \item Design of two power-efficient and high output gain class-E PAs with single and two-stage configurations.
  \item Fabrication of the proposed single stage and 2-stage PA designs using the Taconic printed circuit board (PCB).
  \item Comparison of simulation as well as experimental results for output gain and power added efficiency (PAE) for both the proposed class-E PA.
\end{itemize}
The rest of the paper is organized as follows, Section II explains the design methodology with class-E configuration. Section II also provide simulation and fabricated design and results for single stage class-E PA. Section III provides simulation and fabricated design and results for 2-stage class-E PA. Section IV provides a detailed comparison of the results of single stage and 2-stage class-E PA which are designed in the earlier sections. Finally section V concludes the paper.
\section{Design Methodology with Class-E Configuration}
Fig. 2 shows an amplifier with dual harmonic suppression configuration which helps to enhance its efficiency. The input impedance matching is provided by the capacitors $C_2$ and $C_3$ between the source and gate of the transistor, while the output impedance matching is achieved with the help of the capacitor $C_6$ and matching stubs. Capacitors $C_1$ and $C_7$ work as the DC-Block capacitors while the inductors $L_1$, $L_4$, and $L_5$ prevents any alternating current (AC) signal from reaching the direct current (DC) supply. Maximum gain from the amplifier is derived by optimizing the input and output impedance matching circuits. DC supply voltage is another important factor which greatly affects the gain and efficiency so, proper biasing is provided by the use of resistors at the gate and drain of the transistor.
%\begin{table}[]
%\centering
%\caption{Measured stability factor of transistor ATF-53189}
%\label{my-label}
%\begin{tabular}{|c|c|}
%\hline
%\textbf{Frequency} & \textbf{StabMeas-1} \\ \hline
%800.0 MHz          & 1.700               \\ \hline
%1.000 GHz          & 1.702               \\ \hline
%1.200 GHz          & 1.703               \\ \hline
%1.400 GHz          & 1.706               \\ \hline
%1.600 GHz          & 1.708               \\ \hline
%1.800 GHz          & 1.711               \\ \hline
%2.000 GHz          & 1.717               \\ \hline
%2.200 GHz          & 1.724               \\ \hline
%2.400 GHz          & 1.732               \\ \hline
%2.600 GHz          & 1.741               \\ \hline
%2.800 GHz          & 1.752               \\ \hline
%3.000 GHz          & 1.764               \\ \hline
%\end{tabular}
%\end{table}

In an ideal Class-E PA, a square wave is used in order to drive the gate of the transistor, but for shaping the input waveform to square wave, a non-linear gate capacitor cannot be used \cite{banerjee2013adaptive}. For this reason, the harmonic suppression circuits are used which help to shape the waveform and allow only the first harmonic of the frequency of operation. This circuit is placed perpendicular to $C_2$ and $C_3$ because it helps to minimize magnetic coupling between the components. A resistor may be added at the input or output circuit if the transistor is potentially unstable for the desired frequency.
\subsection{Implementation of Proposed Design}
When designing a Class-E PA the selection of transistor is very important, this will define the maximum gain and linearity which we can achieve. If the selected transistor is potentially unstable at the frequency of operation then a resistor will be required at the input or output circuit in order to make the transistor stable but it will reduce the overall output gain of the P.A, that's why a transistor with stable characteristic at the desired frequency is preferred. After selection of the transistor, it is best practice to test its linear model on the software which should produce the same results of current and voltage as mentioned in its datasheet. In the proposed PA we have used ATF-53189 transistor because of its high efficiency.
\subsection{Simulation Design of Proposed Single Stage P.A.}
The simulation for the design of proposed single stage Class-E PA is performed on advanced design system (ADS). As it is required to find the value of maximum PAE so a harmonic balance should be introduced in this simulation and for that the non linear model of transistor ATF-53189 is used. The main parameters which are needed in order to find the value of PAE are input RF power, output RF power, input DC voltage and input DC current. Parameter sweep range is selected from -10dB up to 20dB.
\begin{figure*}[ht]
\begin{center}
	\includegraphics[width=7.5in]{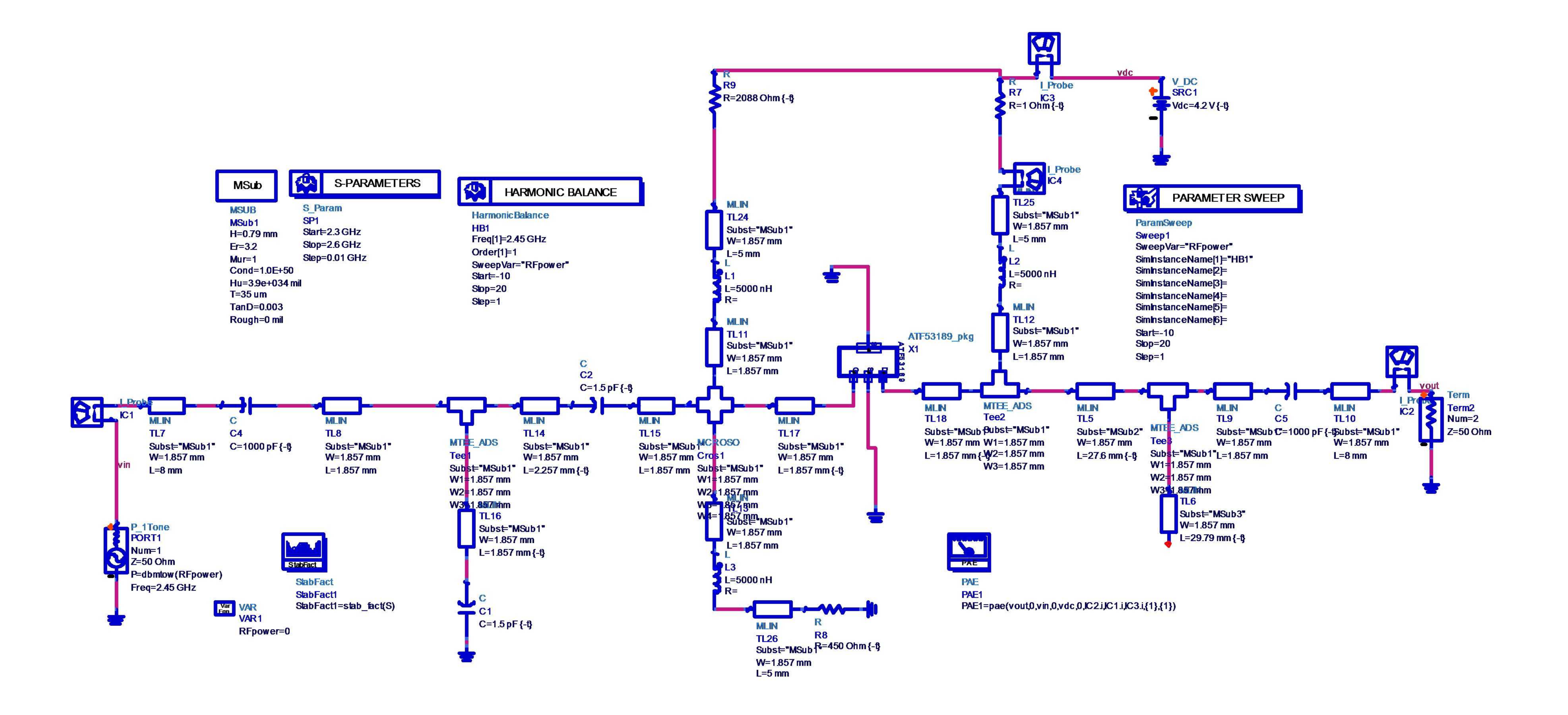}
\caption{Schematic diagram of single-stage PA with input and output matching circuits.}
\label{fig:sysmodel}
\end{center}
\end{figure*}

Fig.3 shows the schematic diagram of PA with input and output matching circuits. The perfectly matched input and output impedance is achieved by tuning the capacitors and transmission lines for maximum output gain. Both the input and output matching circuits are matched with 50 Ohm impedance port.
\begin{figure}
\centering
\subfloat[Output Gain]
{
    \includegraphics[width=1.6in]{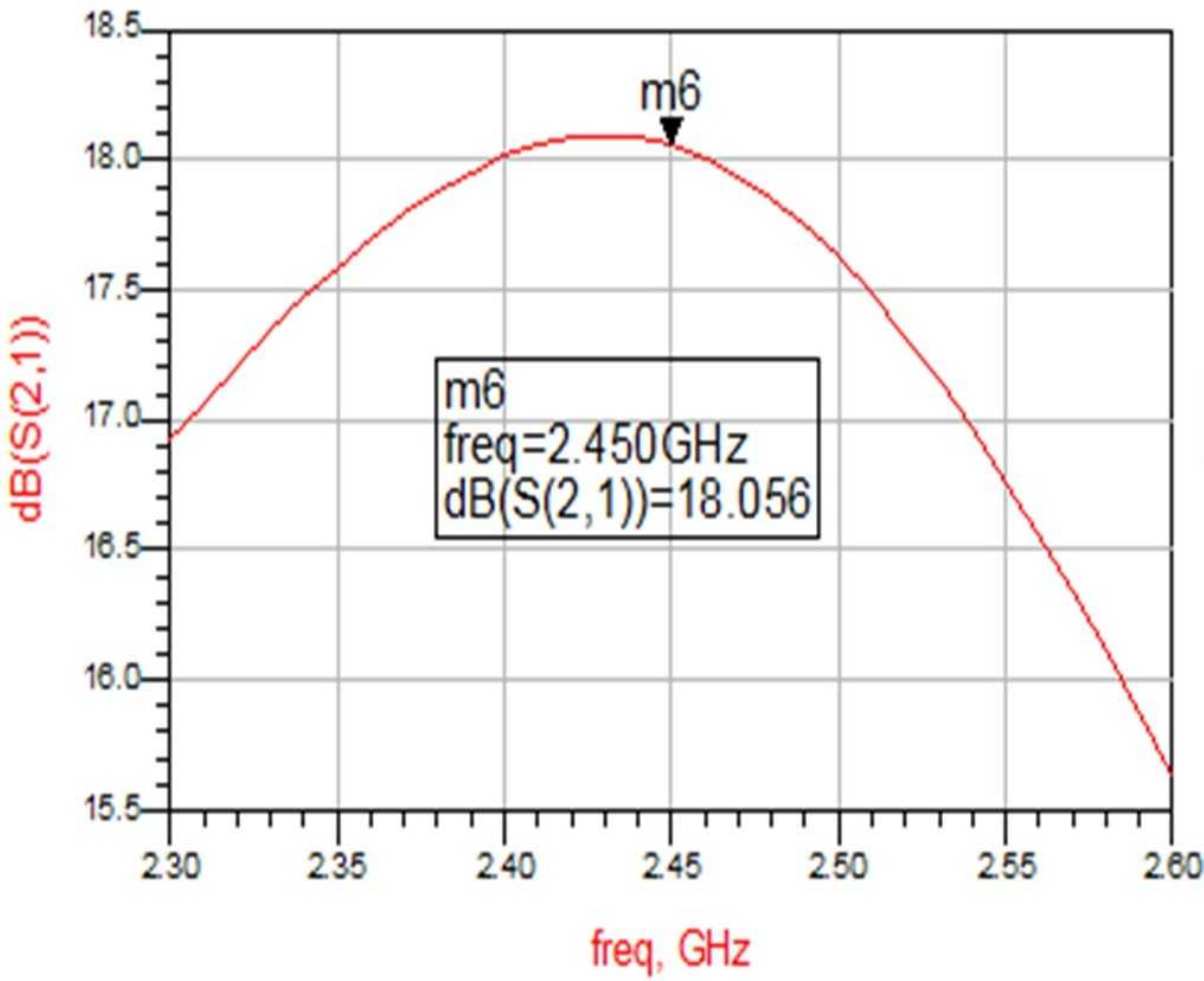}
    \label{fig:foo-1}
}
\subfloat[Power Added Efficiency]
{
    \includegraphics[width=1.6in]{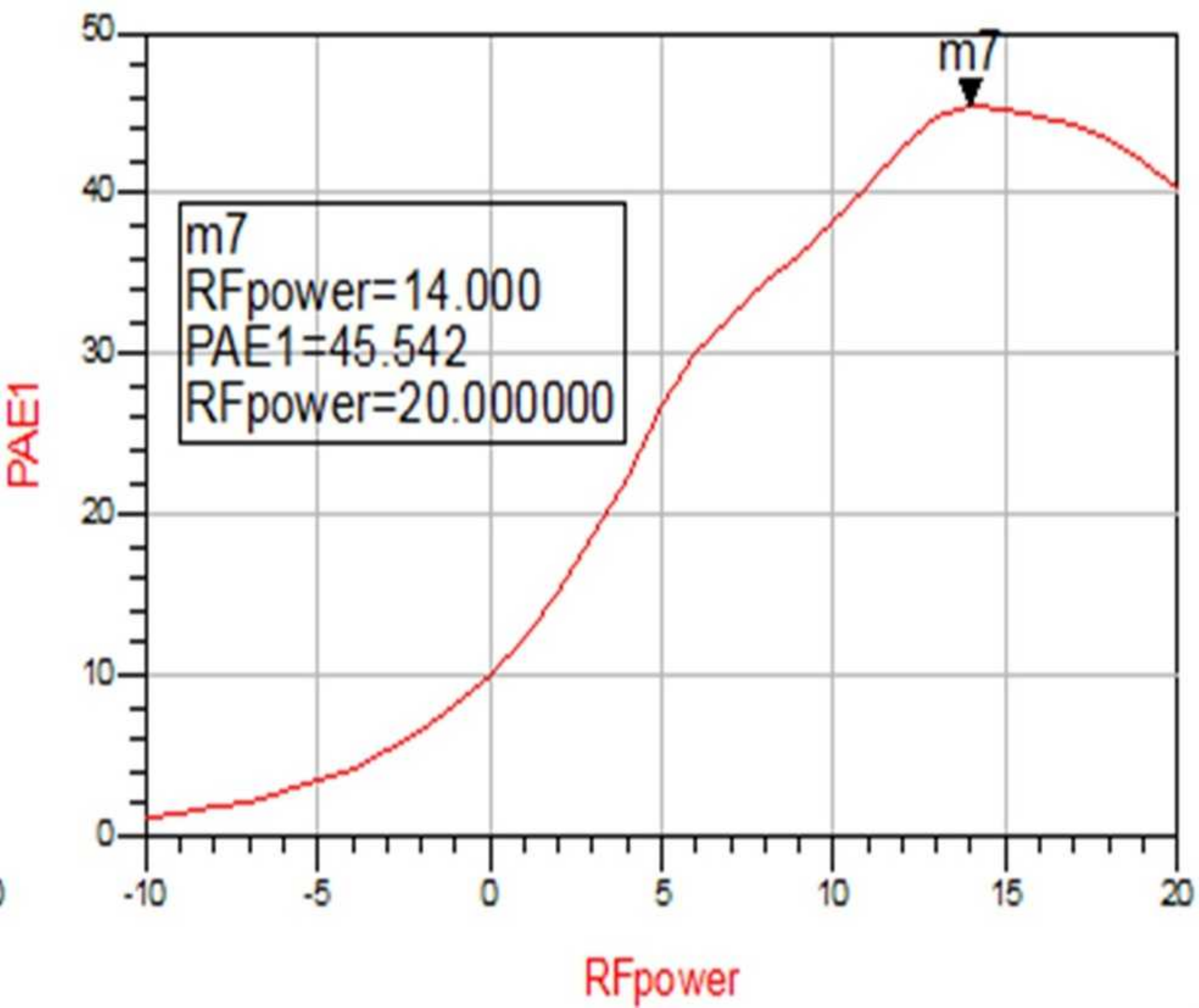}
    \label{fig:foo-2}
}
\caption{Proposed single-stage PA with input and output matching circuits.}
\label{fig:foo}
\end{figure}
Test results for the schematic shown in Fig. 3 are revealed in Fig. 4(a) and 4(b). In Fig. 4(a) the output gain is found to be 18dBm. In Fig. 4(b) the value of total PAE is 45.5\% and it can be seen that the PAE is improved and its maximum value is obtained at an input RF power of 14dBm.
\begin{figure*}[ht]
\begin{center}
	\includegraphics[width=7.5in]{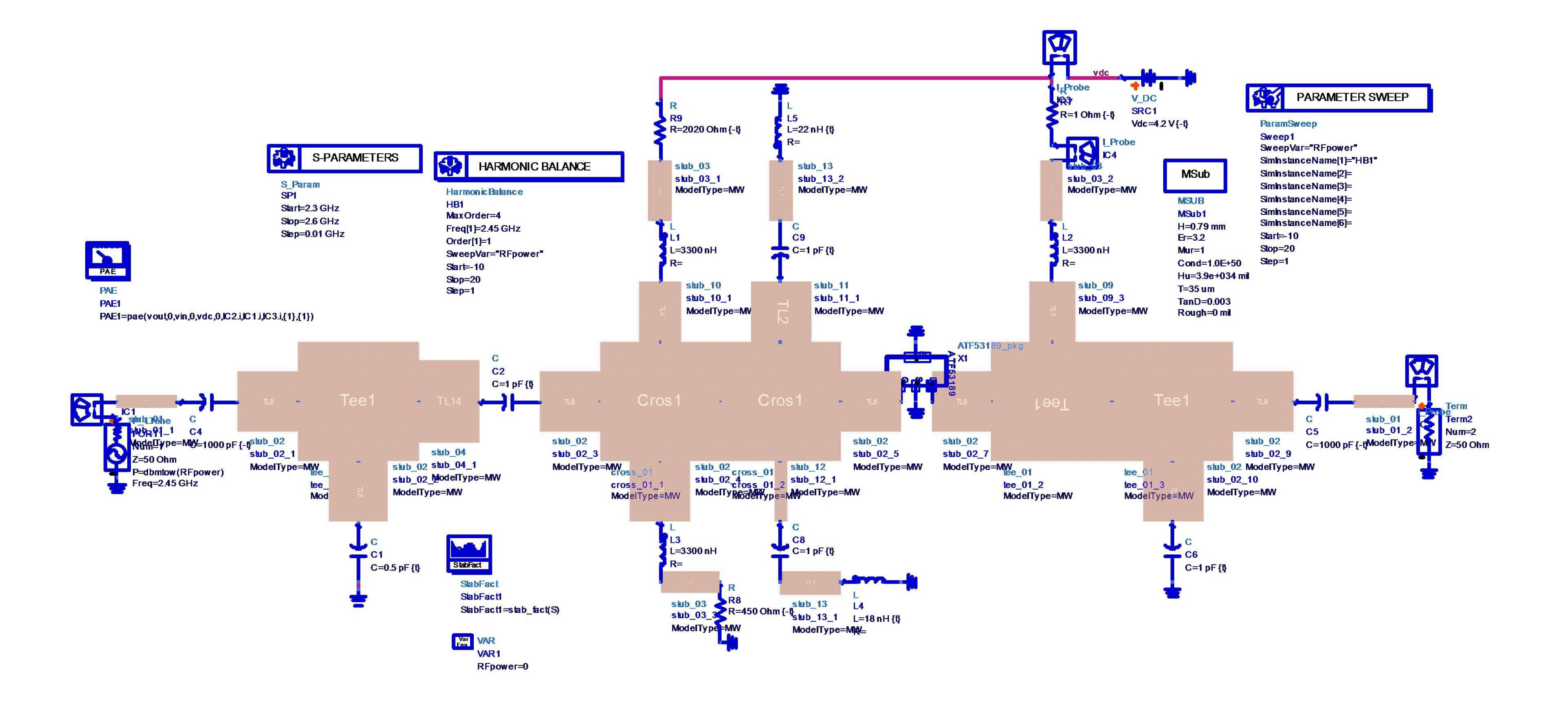}
\caption{Co-simulation of proposed single-stage PA with Harmonic Suppression circuits.}
\label{fig:sysmodel}
\end{center}
\end{figure*}
After the addition of harmonic suppression circuits and optimizing the impedance matching circuits the maximum PAE that could be achieved is 53\% at an input RF power of 16dBm while appending the H.S circuits the new gain of P.A is now 17.5dBm. Fig. 6(a) and 6(b) shows the final S-Parameter results together with PAE.
\begin{figure}
\centering
\subfloat[Co-simulated S-parameters]
{
    \includegraphics[width=1.6in]{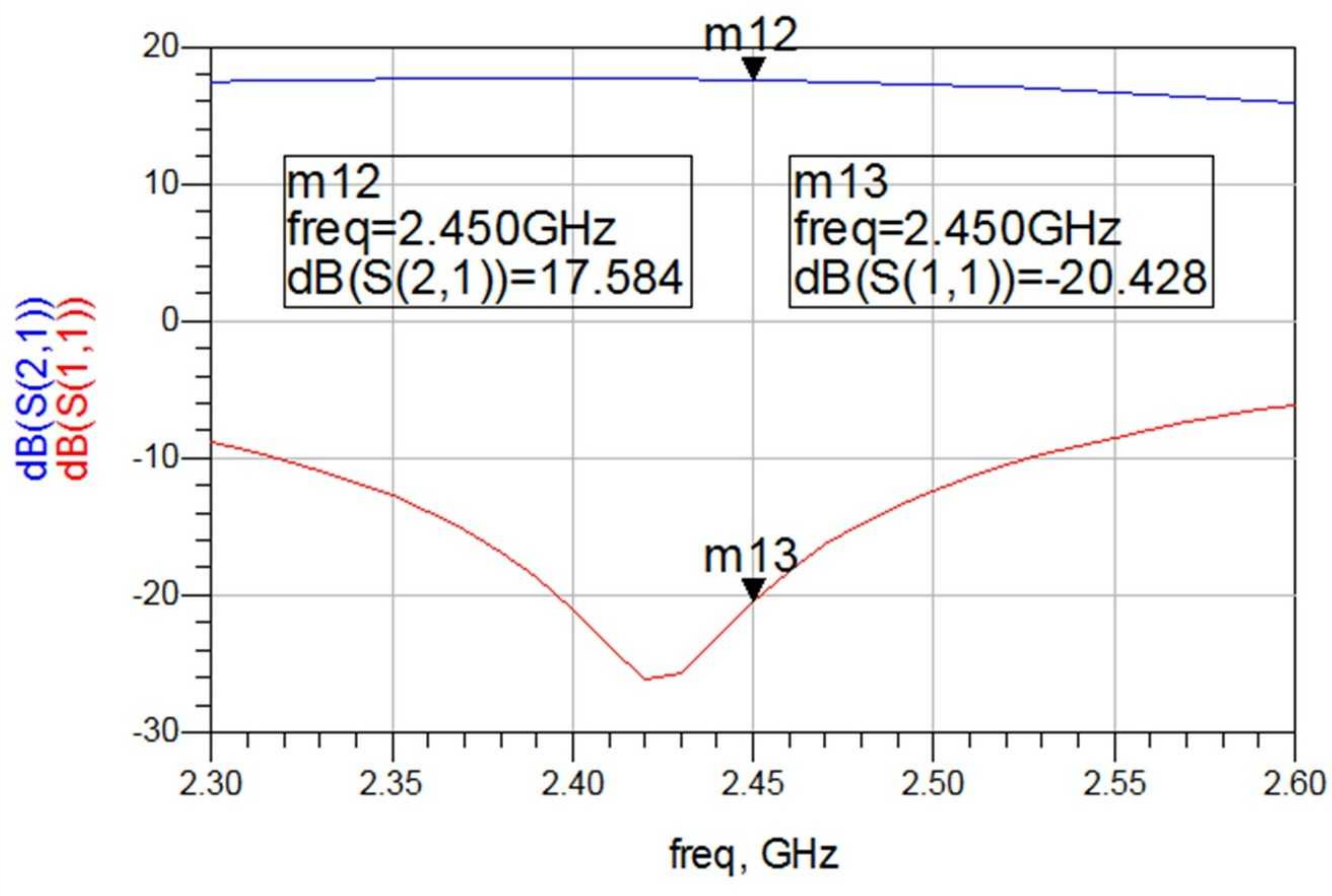}
    \label{fig:foo-1}
}
\subfloat[Co-simulated PAE]
{
    \includegraphics[width=1.6in]{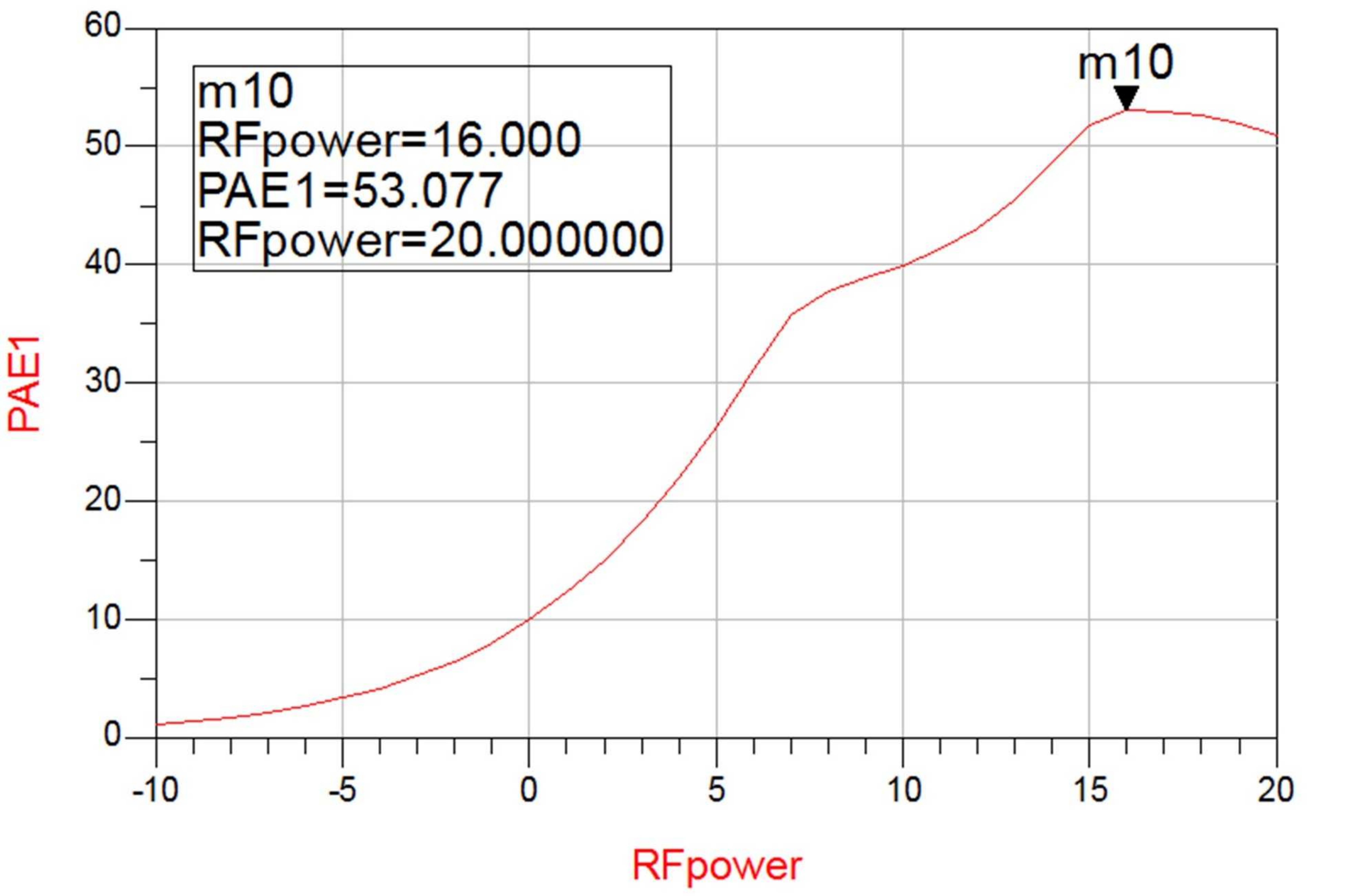}
    \label{fig:foo-2}
}
\caption{Proposed single-stage class-E PA with Harmonic Suppression circuits.}
\label{fig:foo}
\end{figure}

To further improve the PAE of this Class-E PA two harmonic suppression circuits are added at the input side of amplifier. These circuits help to reduce the power consumed at the harmonic frequencies which will enhance the overall efficiency of this amplifier. The co-simulation of the added harmonic suppression circuits and optimized input output impedance matching circuits is shown in Fig. 5.
\begin{figure}[ht]
\begin{center}
	\includegraphics[width=3in]{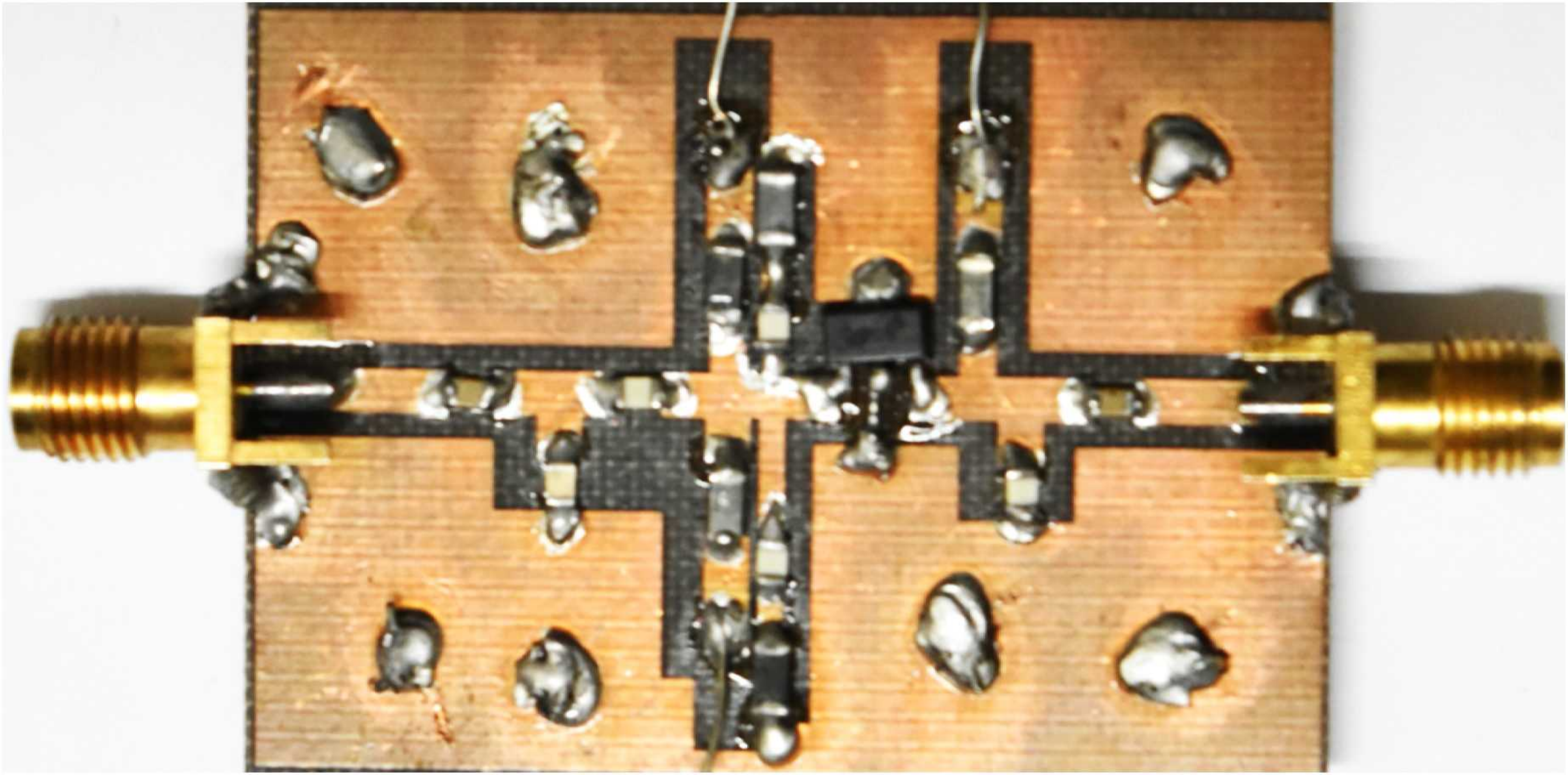}
\caption{Fabricated design of proposed single-stage class-E PA.}
\label{fig:sysmodel}
\end{center}
\end{figure}

\begin{figure}
\centering
\subfloat[Measured Output Gain]
{
    \includegraphics[width=1.6in]{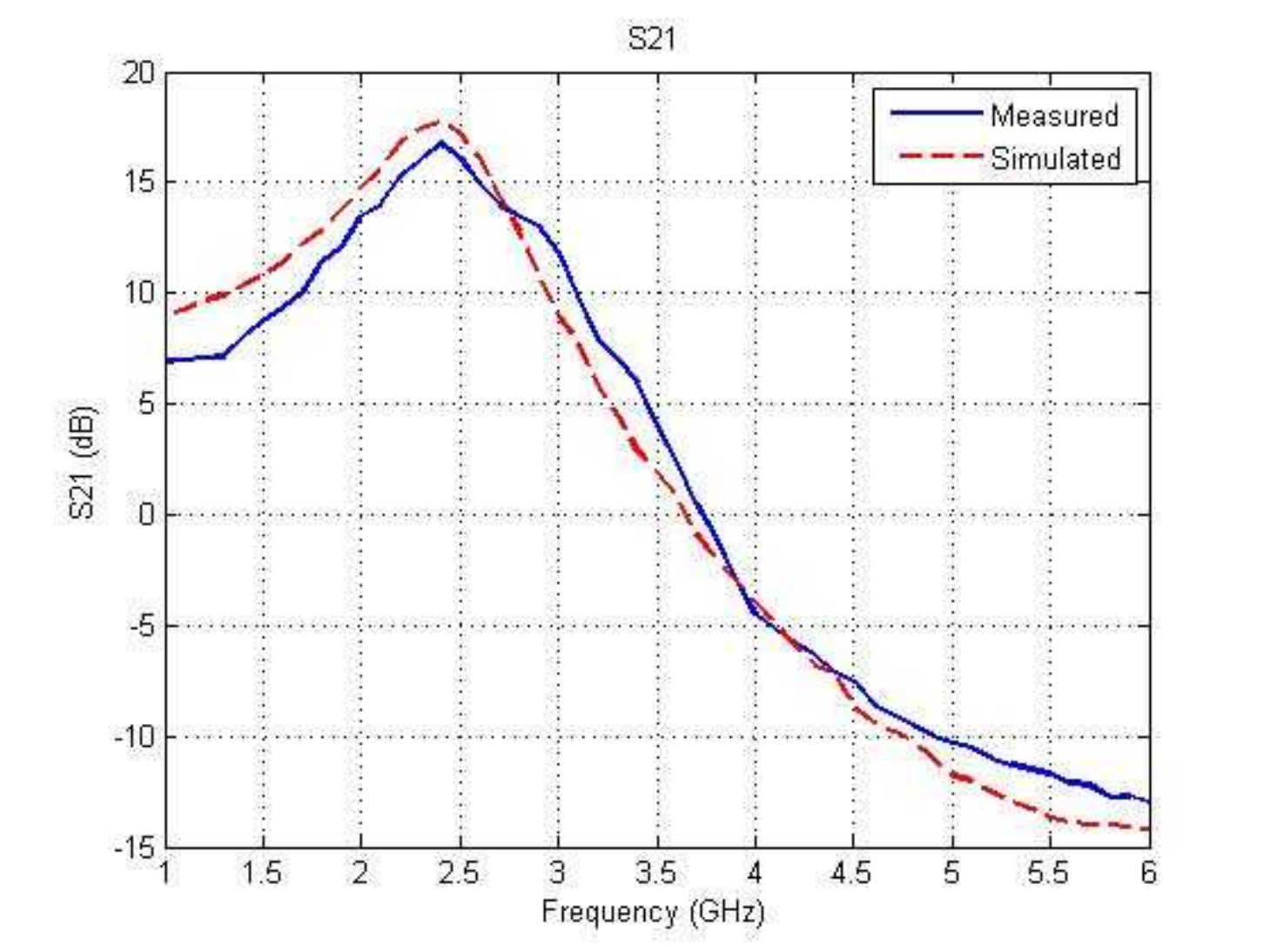}
    \label{fig:foo-1}
}
\subfloat[Measured PAE]
{
    \includegraphics[width=1.6in]{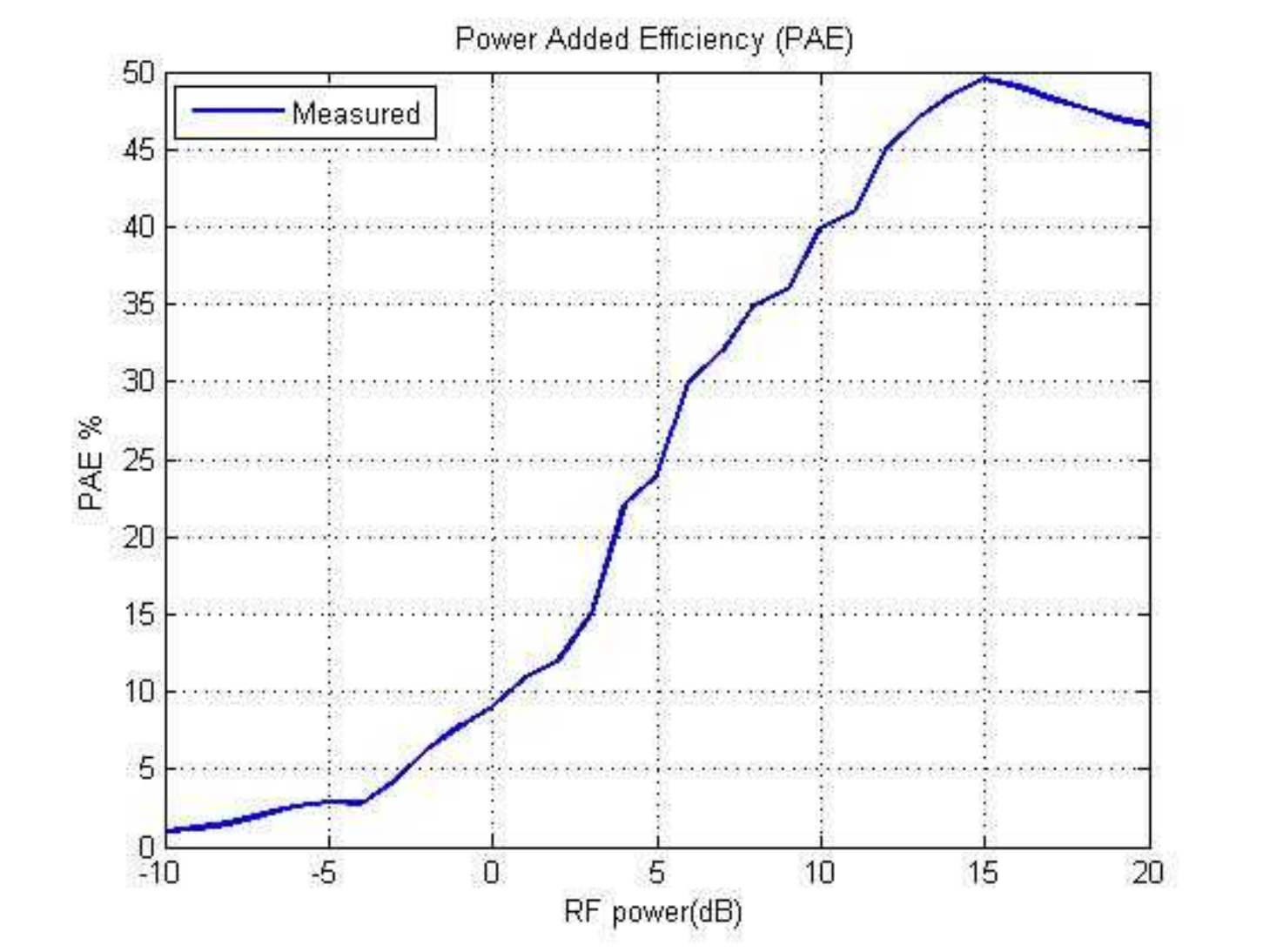}
    \label{fig:foo-2}
}
\caption{Proposed single-stage class-E PA.}
\label{fig:foo}
\end{figure}
\subsection{Fabrication of the Proposed Design}
The proposed single stage PA is fabricated on taconic PCB with thickness of 0.79mm and Dielectric constant 3.2. The size of the PCB is 4.6cm x 3.4cm. The PA is matched to 50 ohm source and load ports. Proper grounding of top and bottom layer is performed by drilling holes and then using a wire to connect the top and bottom layer at different locations on the PCB. Fig. 7 shows the fabricated design of the proposed single stage PA.
\begin{figure*}[ht]
\begin{center}
	\includegraphics[width=7.5in]{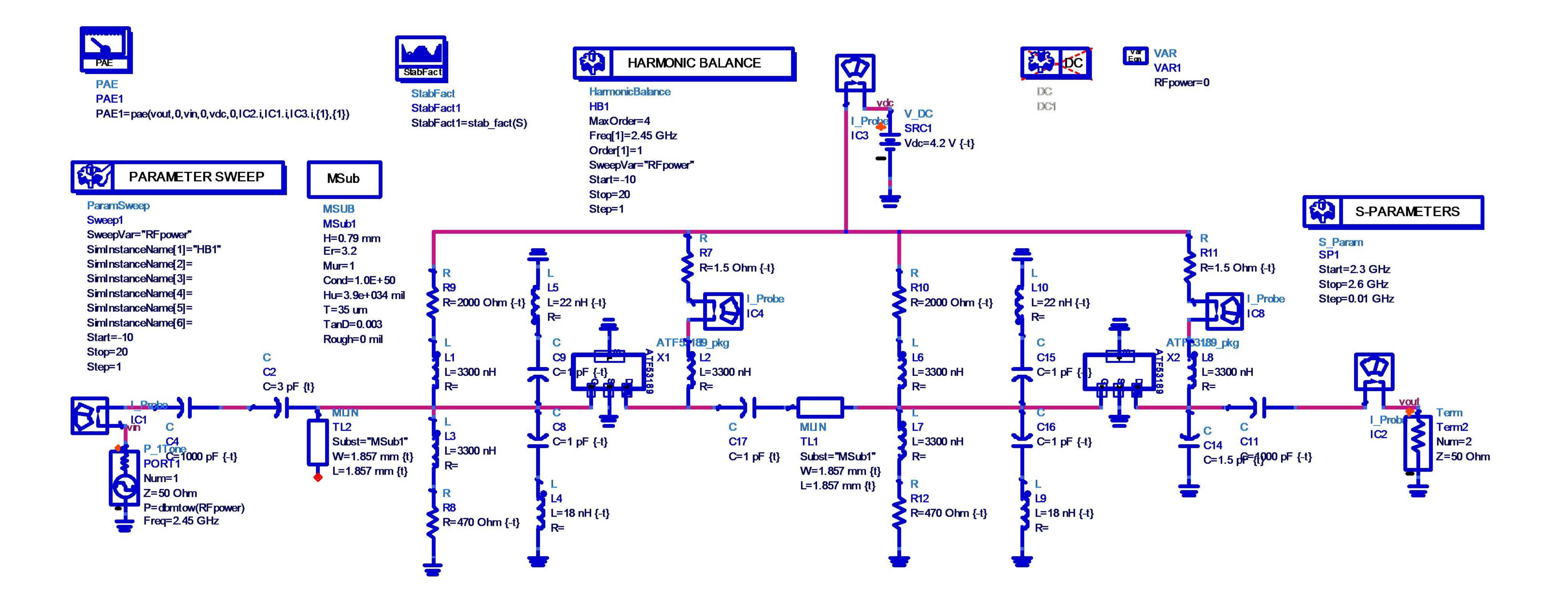}
\caption{Schematic diagram of the proposed 2-stage class-E PA.}
\label{fig:sysmodel}
\end{center}
\end{figure*}
\subsection{Measured Results}
Fig. 8 shows the measured results of the proposed single stage PA where for different values of frequencies, the output gain of the PA is shown in Fig. 8(a). From this figure the measured output gain comes out to be 16.7dB which is very close to the simulated results. The plot for PAE can also be seen in Fig. 8(b). From this graph it can be seen that the maximum efficiency of 49.5\% is achieved at an input RF power of 15dB which is also similar to the simulated results.
\begin{figure}
\centering
\subfloat[$S_{21}$ VS Frequency ]
{
    \includegraphics[width=1.6in]{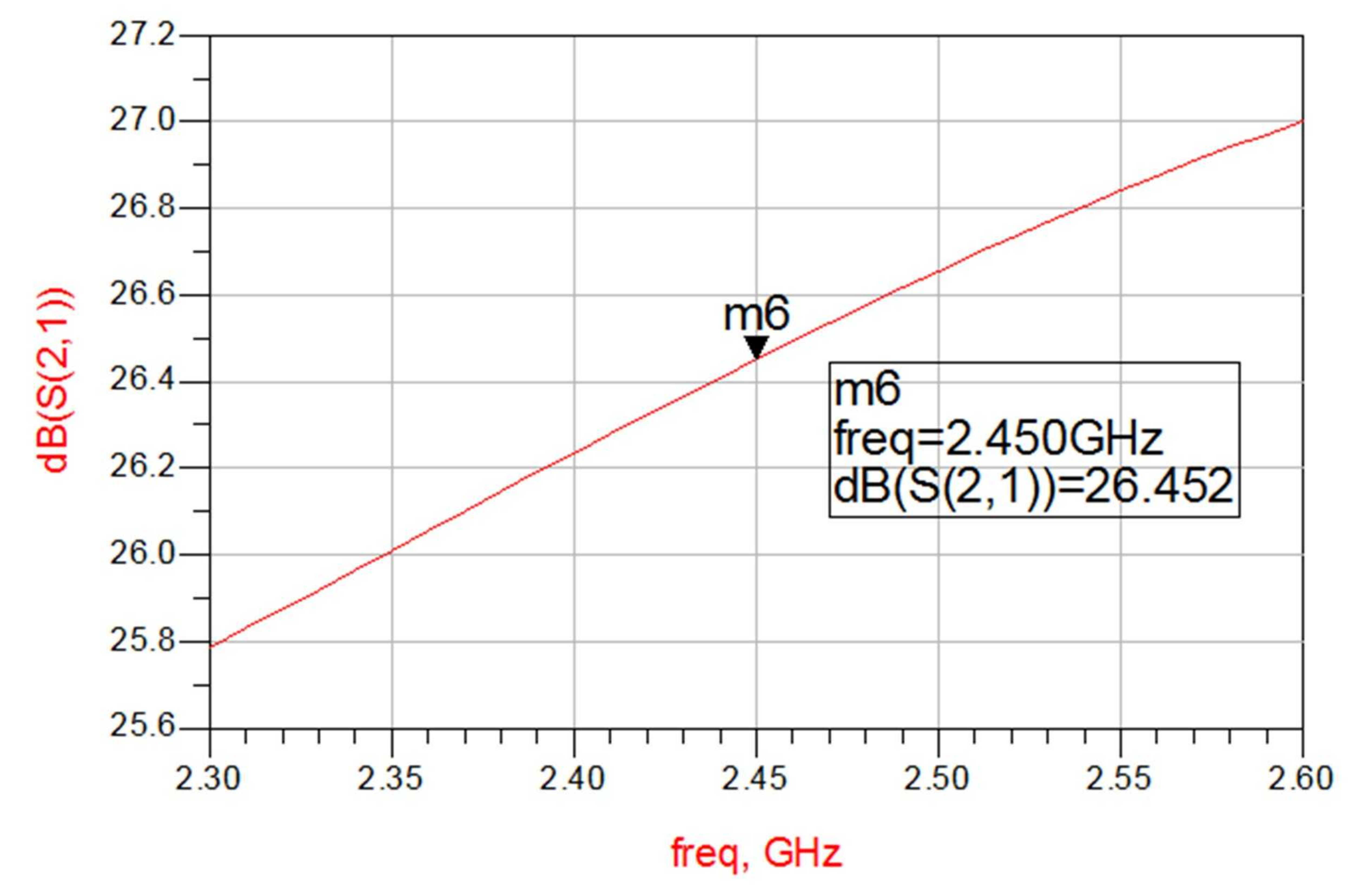}
    \label{fig:foo-1}
}
\subfloat[PAE vs Input RF power]
{
    \includegraphics[width=1.6in]{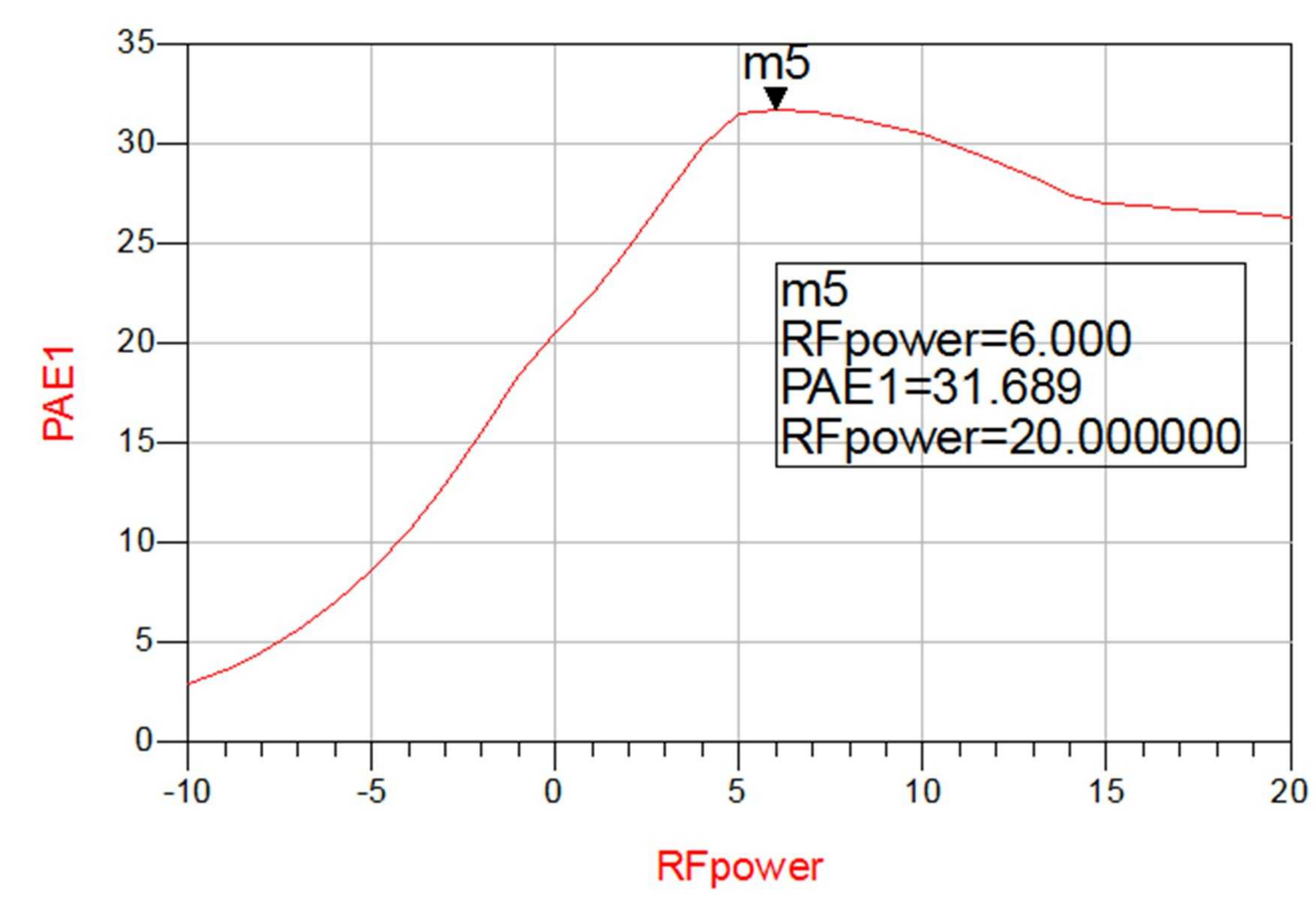}
    \label{fig:foo-2}
}
\caption{Simulated two-stages class-E PA.}
\label{fig:foo}
\end{figure}
\section{Design of Proposed 2-Stage P.A.}
In the proposed 2-stages PA, two single stage PAs are connected in cascade fashion as shown in Fig. 9. The design of 2-stage PA is the same as was for the single stage, the only difference is the impedance matching circuit which is added between the two amplifiers while cascading. Resistors R7, R8 and R9 are the biasing resistors for the 1st stage while resistors R10, R11 and R12 are for 2nd stage of the 2-stages PA. Capacitor C2 and transmission line TL2 are the input impedance matching circuits while capacitor C14 and C11 are the output matching circuits. The impedance matching between the 1st stage and 2nd stage of the Amplifier is achieved with the help of capacitor C17 and transmission line TL1. The capacitor C17 also helps to prevent any interference of 1st stage biasing circuit with the 2nd stage biasing circuit. Capacitors C4 and C11 are the coupling capacitors while inductors L1-L3 and L6-L8 are DC pass inductors to stop the AC signal from going towards DC power supply. Unwanted harmonics are filtered by inductors L4, L5, L9, L10 and capacitors C8, C9, C15 and C16. The Supply voltage VCC is kept 4.2 volts as was for single stage PA but its current consumption would be twice than single stage as there are two PA cascaded. This PA design was targeted towards better output gain while compensating for little loss in efficiency.
\begin{figure}
\centering
\subfloat[$S_{21}$ VS Frequency ]
{
    \includegraphics[width=1.6in]{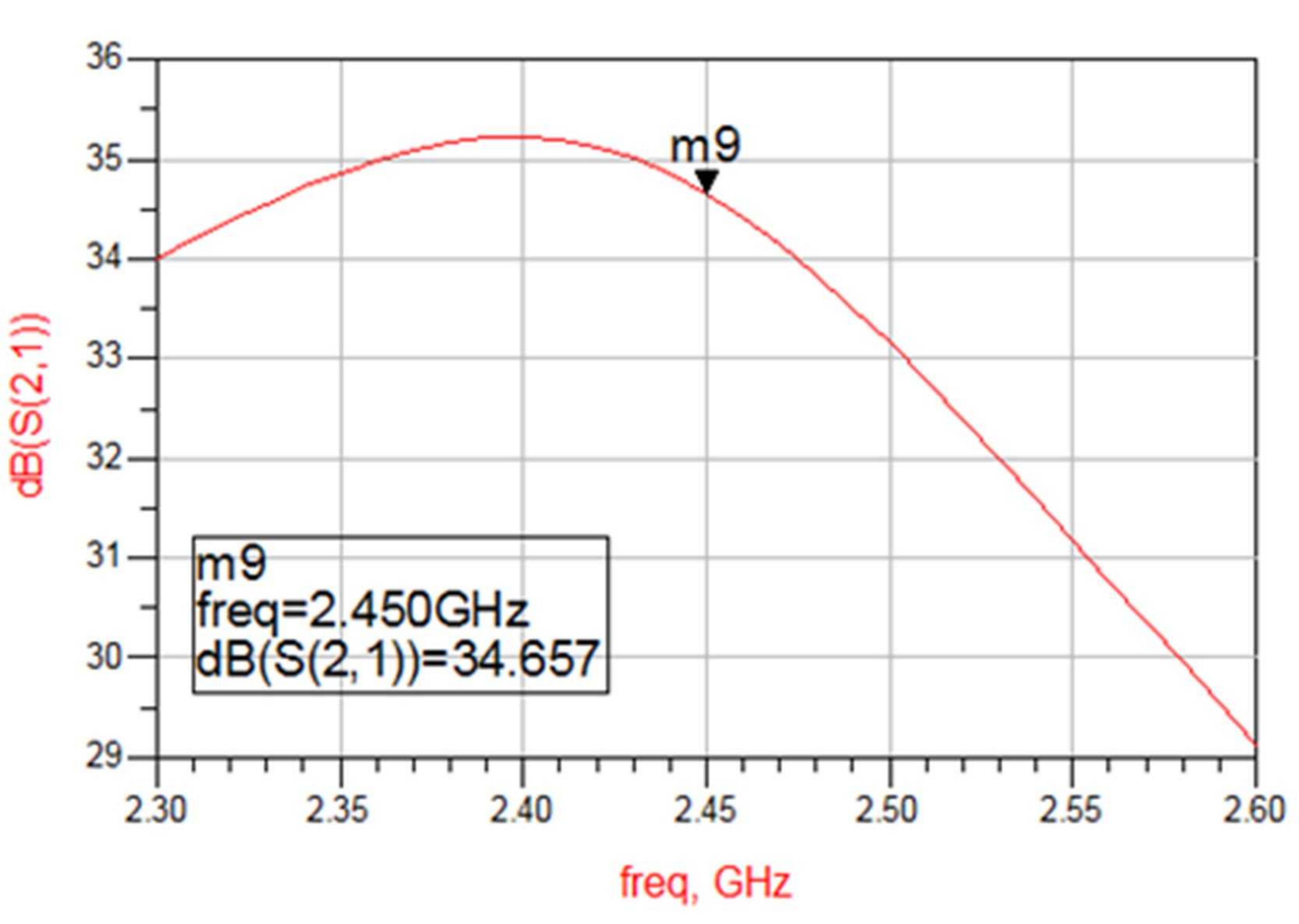}
    \label{fig:foo-1}
}
\subfloat[PAE vs Input RF power]
{
    \includegraphics[width=1.6in]{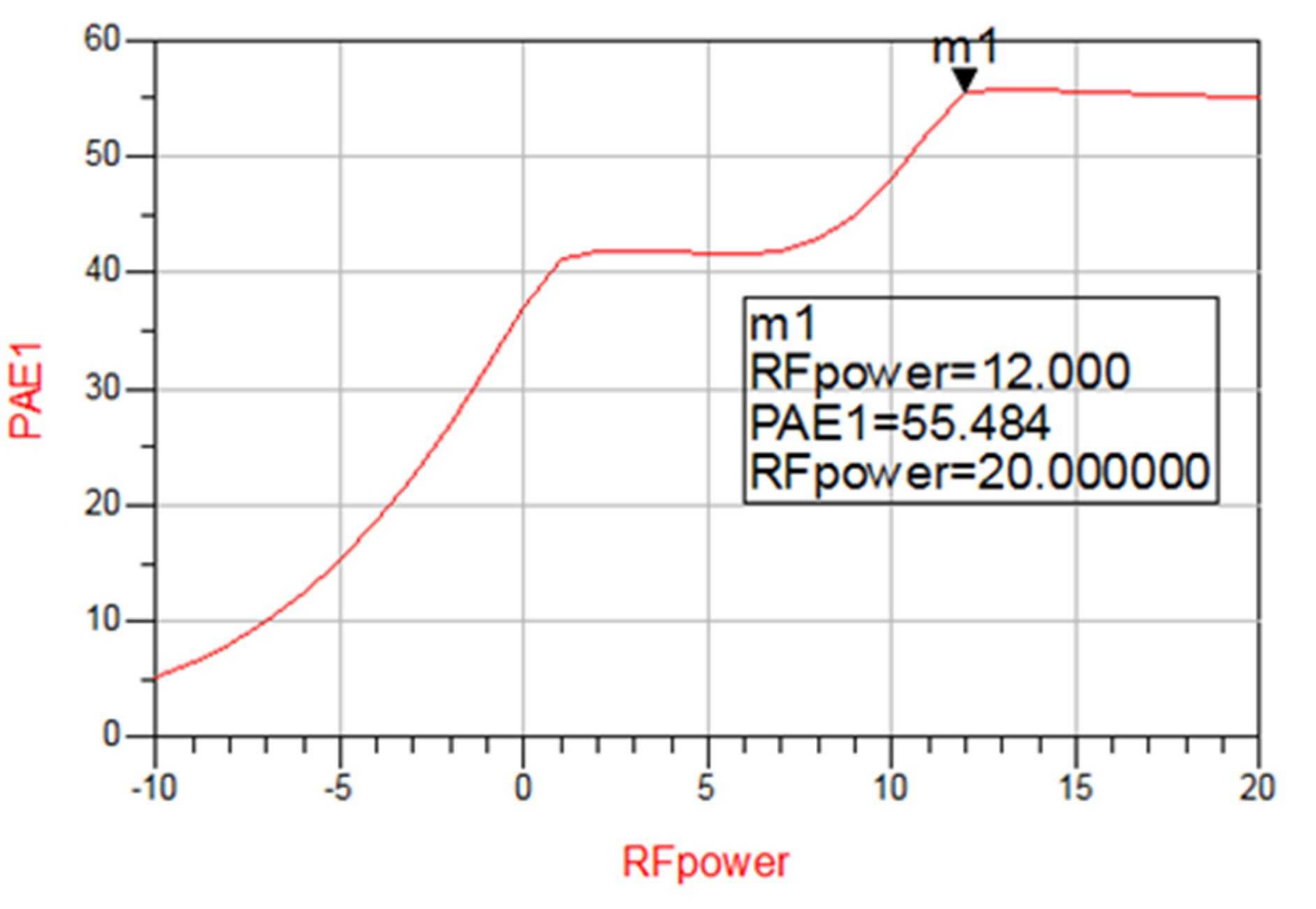}
    \label{fig:foo-2}
}
\caption{Co-simulated two-stages class-E PA.}
\label{fig:foo}
\end{figure}
From Fig. 10(a) and 10(b), the output gain parameter S21 comes out to be 26.45dB while the PAE is 31.6\%. For optimizing and tuning the two-stages PA, first the schematic diagram is moved for co-simulation and then it is simulated to get the optimized results, next step is the tuning where different components are selected for tuning and while changing the values of lumped components results are improved. In our design the input-output matching circuits and DC blocking capacitors are tuned to get maximum gain from this 2-stages PA.
\begin{figure}[ht]
\begin{center}
	\includegraphics[width=3in]{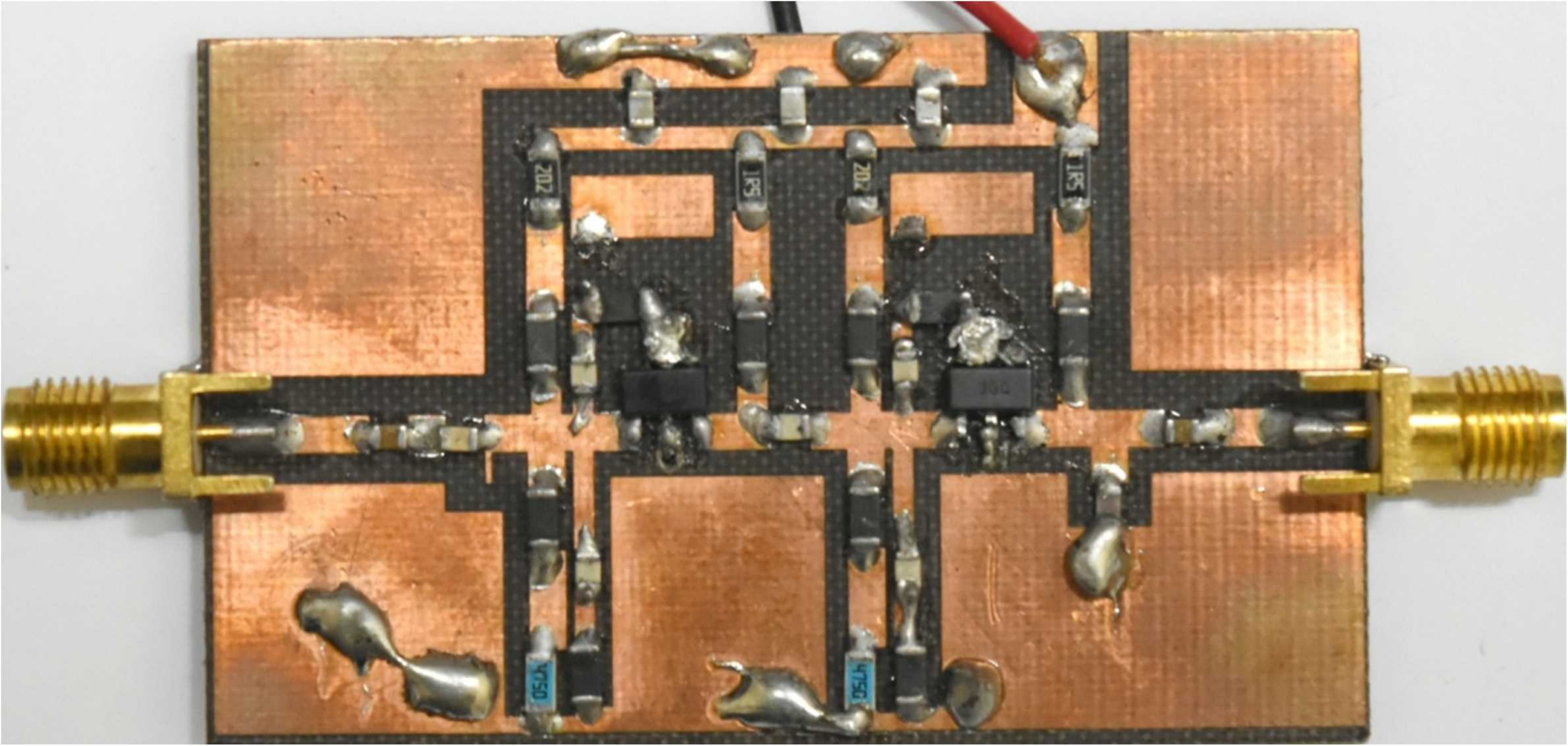}
\caption{Fabricated design of the Proposed 2-stages class-E PA.}
\label{fig:sysmodel}
\end{center}
\end{figure}
In Fig. 11(a) and 11(b) plots for gain and PAE are shown. It is clear that the optimizing and tuning of schematic in co-simulation has increased the gain of PA and now it has improved up to 34.6dB which is more than twice as was for single stage PA. The PAE came out to be 55.4\%.
\begin{figure}
\centering
\subfloat[Measured Output Gain ]
{
    \includegraphics[width=1.6in]{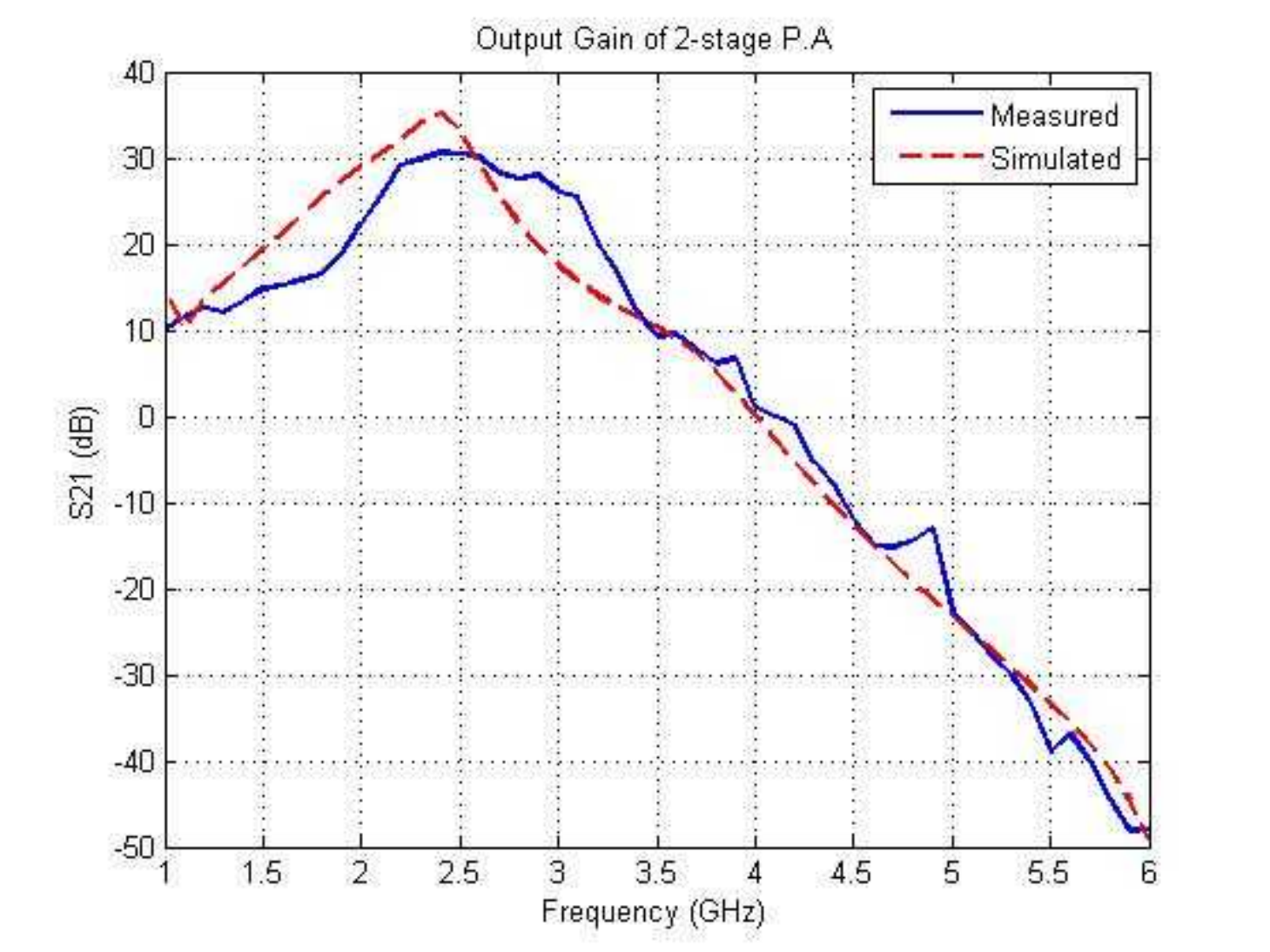}
    \label{fig:foo-1}
}
\subfloat[Measured PAE]
{
    \includegraphics[width=1.6in]{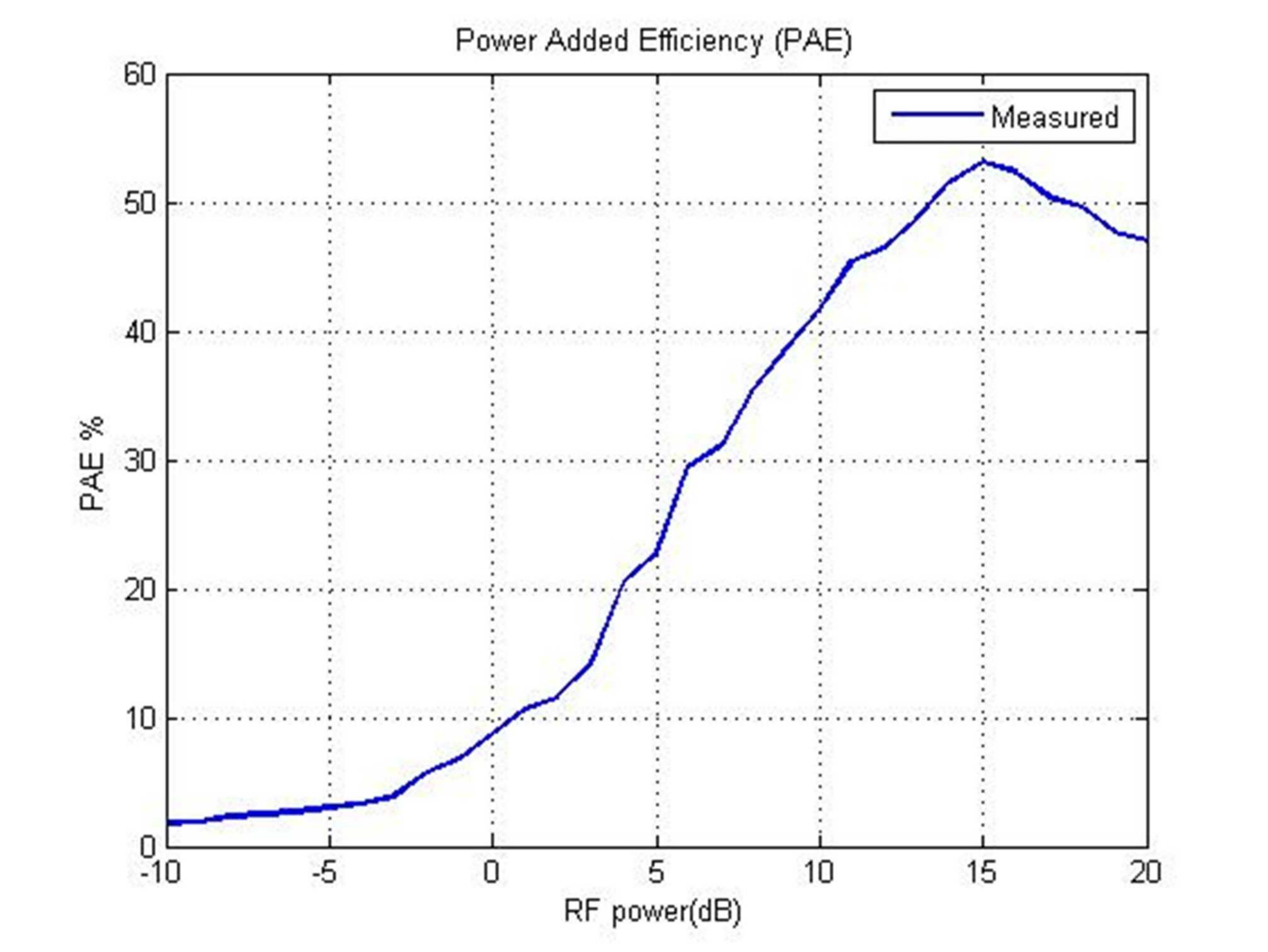}
    \label{fig:foo-2}
}
\caption{Measured results of the proposed two-stages class-E PA.}
\label{fig:foo}
\end{figure}
\subsection{Fabrication of the Proposed Design}
The 2-stages PA is fabricated on taconic PCB with dielectric constant value of 3.2 and thickness of 0.79mm as shown in Fig. 12. The dimensions of the capacitors and inductors used here are according to the standard of 0805 and 1206 respectively. The grounding of transistor is directly done using thin wires with the lower side of the PCB as it affects the gain of the PA. The dimensions of the PCB are 5.9cm x 3.6cm.
\subsection{Measured Results}
The measured output gain of 2-stage PA is shown in Fig.13(a). The maximum gain achieved at 2.4GHz is 30.5dB which is very close to the simulated results. For 2-stage PA the maximum PAE is measured and found to be 53.1\% which is shown in Fig.13(b).
\section{Result Comparison of Single Stage and 2-Stage P.A.}
\begin{figure}
\centering
\subfloat[Measured Output Gain]
{
    \includegraphics[width=1.6in]{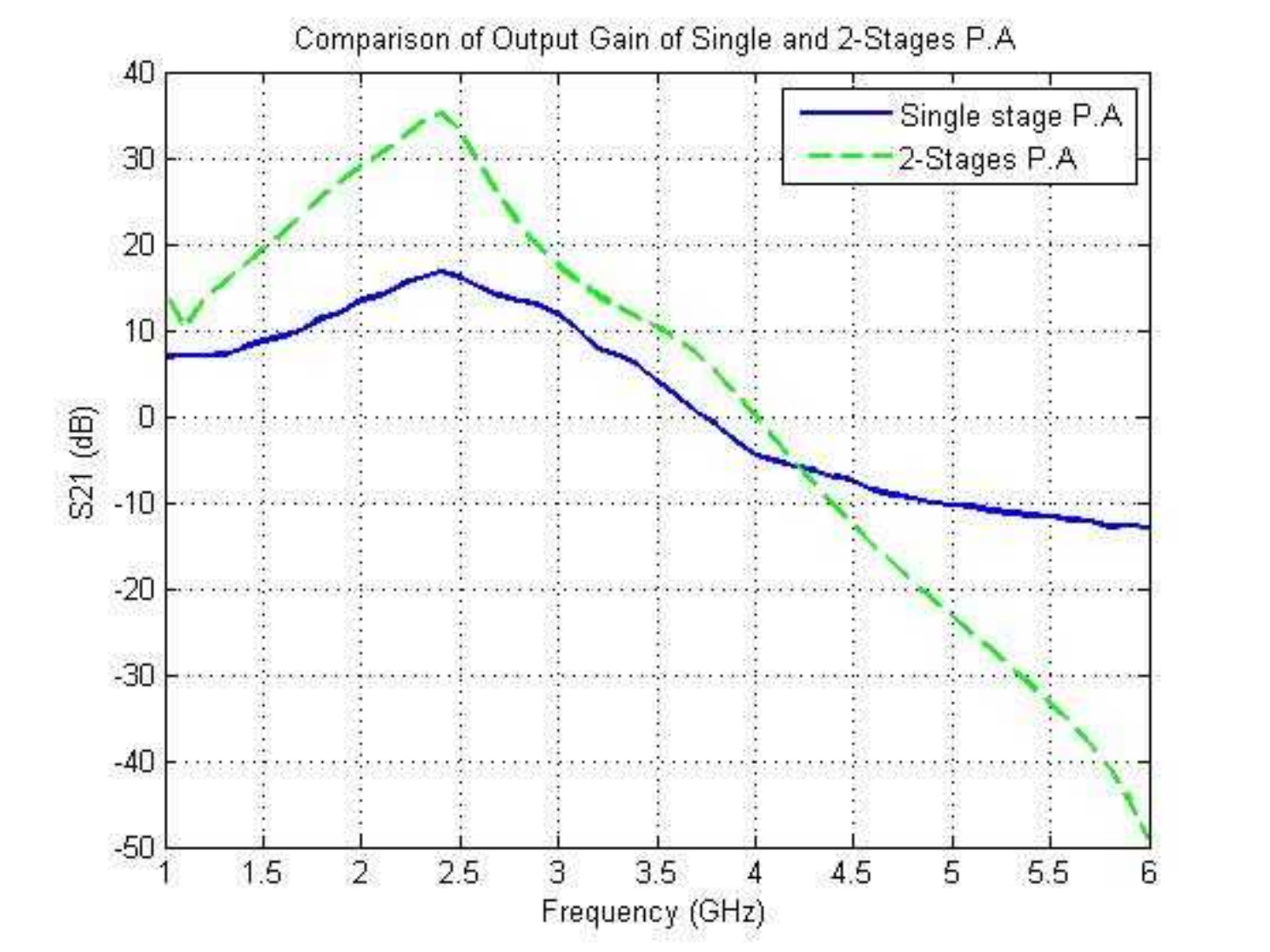}
    \label{fig:foo-1}
}
\subfloat[Measured PAE]
{
    \includegraphics[width=1.6in]{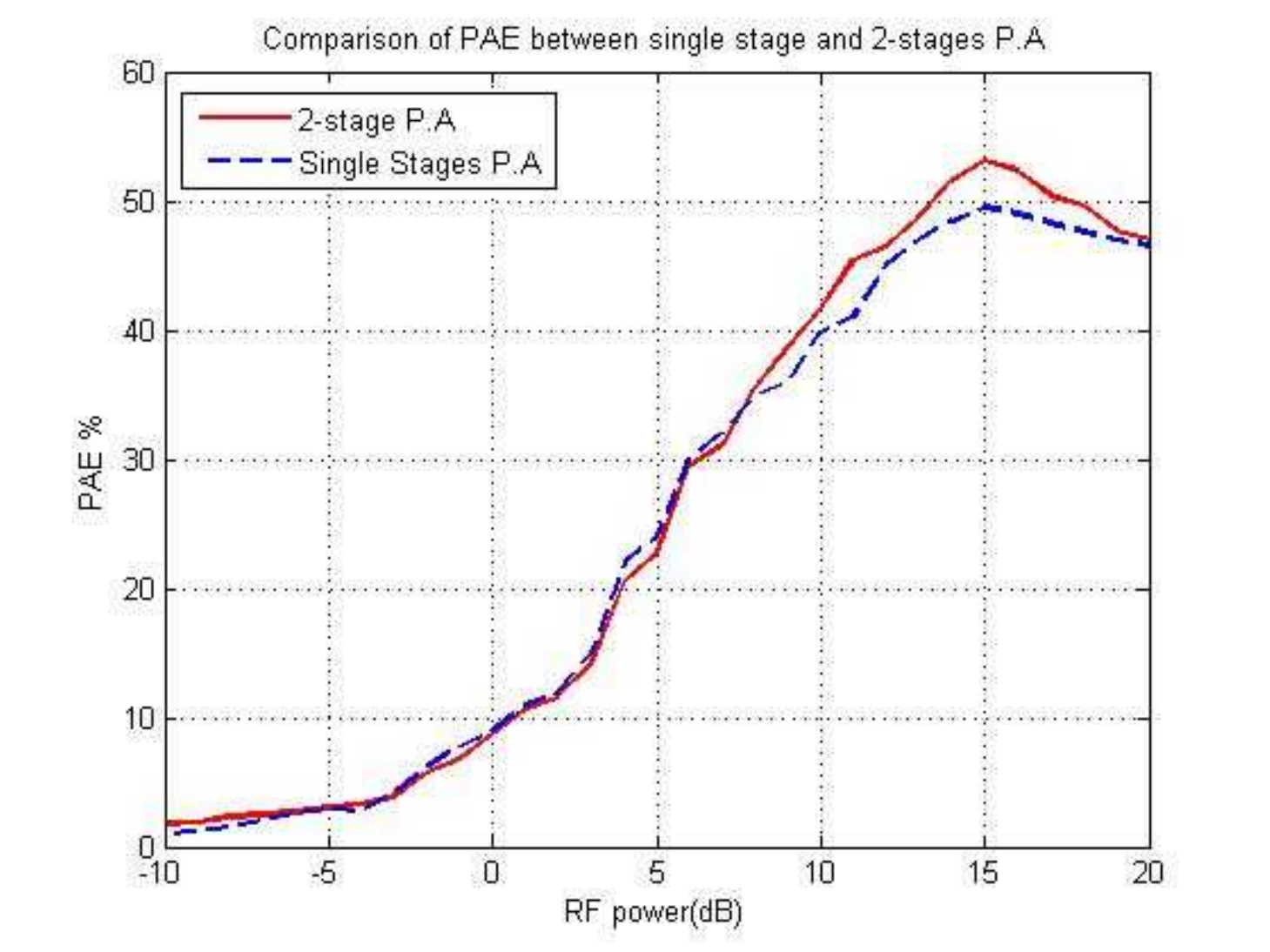}
    \label{fig:foo-2}
}
\caption{Comparison between single-stage and 2-stage class-E PAs.}
\label{fig:foo}
\end{figure}
A comparison of measured total output gain and PAE is carried out between the single stage and 2-stages PA and is shown in Fig.14(a) and 14(b). As can be seen from Fig.14(a) the output Gain of 2-stages PA is twofold as compared to the gain of single stage PA but the graph is more linear incase of single stage. In Fig.14(b) there is a linear rise of PAE for both single stage and 2-stages PA but the later one is showing a greater peak as compared to the single stage.

One can conclude further from the above presented results that the proposed 2-stage PA is far more better than the single stage for total output gain as well as PAE, but if the requirement is just the linearity than the proposed single stage PA would be a better choice. Since the output gain of 2-stage PA is twice than the single stage so for a longer distance communication this amplifier would be an appropriate choice, also it will give better efficiency but it will cost more input DC power whereas if there is a requirement of security and privacy then the proposed single stage PA should be selected as it will transmit signals to a shorter distance and there will be less chances for any intruder to hack the signal.
\section{Conclusion}
In this work, the design of an efficient single stage and 2-stage PAs have been presented with high output gain. In order to improve the efficiency and output gain of these PAs, harmonic suppression and the optimized impedance matching circuits are used.

The proposed PAs shows good simulation and experimental results, where the measured efficiency of single stage and 2-stage PAs comes out to be $49.5\%$ and $51.3\%$, respectively. Similarly the experimental output gain for single stage and 2-stages PAs are $16.7 dB$ and $30.5 dB$, respectively. These results show that the proposed PAs are very useful in various IoT applications which are energy and resource-constrained. Applications like smart farming and connected agriculture require such devices which are power-efficient and our designed single stage class-E PA can be a good candidate for it. On the other hand, for applications with high-efficiency requirements such as factory and process automation, our designed 2-stage class-E PA can play an important role.
\appendices
\footnotesize{
\bibliographystyle{IEEEtran}

}

\end{document}